\PassOptionsToPackage{desactivate}{linenoaa}
\documentclass{aa}  

\usepackage[varg]{txfonts}
\usepackage{natbib}
\bibpunct{(}{)}{;}{a}{}{,} 

\usepackage{graphicx}
\usepackage{amsmath, amssymb}
\usepackage[colorlinks=true,
  linkcolor=blue,
  citecolor=blue,
  filecolor=blue,
  urlcolor=blue]{hyperref}
\usepackage{physics}
\usepackage{tikz}
\usetikzlibrary{positioning, fit, arrows.meta}
\usepackage[capitalise]{cleveref}
\crefname{section}{Sect.}{Sects.}
\usepackage{siunitx}
\usepackage{multirow}
\usepackage{comment} 
\usepackage{xcolor} 

\definecolor{beaublue}{rgb}{0.74, 0.83, 0.9}
\definecolor{mustard}{rgb}{1.0, 0.86, 0.35}
\definecolor{asparagus}{rgb}{0.53, 0.66, 0.42}

\newcommand{\robin}{\texttt{ROBIN-PIP}\xspace}
\newcommand{\borg}{\texttt{BORG}\xspace}

\newcommand{\calP}{\mathcal{P}}
\newcommand{\calM}{\mathcal{M}}
\newcommand{\odm}{\delta_{\rm m}}

\newcommand{\nbody}{\(N\)-body\xspace}
\newcommand{\btheta}{\boldsymbol{\theta}}
\newcommand{\data}{\boldsymbol{d}}

\newcommand{\sety}{\mathcal{R}}
\newcommand{\icparam}{\boldsymbol{x}}
\newcommand{\biasparam}{\boldsymbol{\alpha}}

\newcommand{\Mpch}{\ensuremath{h^{-1}\,\mathrm{Mpc}}}
\newcommand{\hMsun}{\ensuremath{h^{-1}\,{\mathrm{M}}_\odot}}

\newcommand{\cmark}{\checkmark}
\newcommand{\xmark}{$\times$}

\begin{document}

\title{\texttt{ROBIN-PIP:} Robust Bayesian Field-Level Inference with Physics-Informed Priors}

  \author{
      Ludvig Doeser\inst{1,2}\fnmsep\thanks{Corresponding author; \email{ludvig.doeser@fysik.su.se}}
      \and
      Simon Ding\inst{3}
      \and
      Guilhem Lavaux\inst{3}
      \and
      Jens Jasche\inst{1}
    }

  \institute{The Oskar Klein Centre, Department of Physics, Stockholm University,
             Albanova University Center, SE 106 91 Stockholm, Sweden
    \and
            Center for Computational Astrophysics, Flatiron Institute, 162 5th Avenue, New York, NY 10010, USA
    \and
    Sorbonne Université, CNRS, UMR 7095, Institut d’Astrophysique de Paris,
             98 bis boulevard Arago, 75014 Paris, France
             }
            
  \date{Received XXX / Accepted YYY}

 
\abstract
{
Accurately extracting information from galaxy surveys, particularly at non-linear scales, requires increasingly complex and flexible forward models of the observed galaxy distribution. A key component is the galaxy bias relation, which connects the underlying matter distribution to observed galaxies and requires more expressive representations than those currently incorporated into Bayesian field-level inference. Certain parameterized models for this mapping can be overparameterized, with individual parameters carrying no direct physical interpretation, making them challenging to constrain.
}
{
Our objective is to guide the inference of flexible model parameters towards physically plausible regions of parameter space through physics-informed priors. These priors are derived from high-fidelity simulations, thereby enabling the indirect use of simulations that cannot be integrated directly into gradient-based field-level inference.
}
{
To this end, we introduce the \robin\ (\textbf{RO}bust \textbf{B}ayesian \textbf{IN}ference with \textbf{P}hysics-\textbf{I}nformed \textbf{P}riors) framework. As a proof of concept, we jointly infer the cosmic initial conditions and the parameters of a truncated power-law galaxy bias model for a simulated universe. We integrate \robin into the Bayesian Origin Reconstruction from Galaxies algorithm and benchmark against inference without the simulation-based prior.
}
{With the additional prior, posterior standard deviations of the bias model parameters are reduced by up to $20\%$ and autocorrelation lengths are reduced from $730$--$870$ to $440$--$550$, increasing the effective sample sizes. We find no bias in the recovered initial conditions, demonstrating that \robin guides inference towards physically consistent solutions without compromising the data likelihood and highlighting its potential to incorporate overparameterized models in field-level inference.}
{}

  \keywords{methods: statistical --
            dark matter --
            galaxies: halos --
            large-scale structure
of Universe --
            early Universe
            }

  \titlerunning{Physics-Informed Priors for Field-Level Inference}
  \authorrunning{L. Doeser \& S. Ding et al.}
  \maketitle


\section{Introduction}
Modern cosmology is transitioning into the era of Stage-IV surveys, including \textit{Euclid} \citep{euclid, euclid2}, DESI \citep{DESICollaboration2016}, LSST \citep{lsst}, PFS \citep{pfs}, the SKA Observatory \citep{ska1, ska2}, the Nancy Grace Roman Space Telescope \citep{wfirst}, SPHEREx \citep{doreCosmologySPHEREXAllSky2014}, and the Simons Observatory \citep{adeSimonsObservatoryScience2019}, with several of these already delivering data and others expected to do so in the near future. These experiments will probe the large-scale structure of the Universe with unprecedented statistical precision and extend into increasingly small-scale and non-linear regimes. Fully exploiting this information requires computational tools capable of predicting structure formation and its observable signatures with sufficient physical fidelity.

Bayesian field-level inference, in which the three-dimensional galaxy distribution is modelled via numerical simulations \citep[e.g.][]{jaschePhysicalBayesianModelling2019,mcalpineManticoreProjectDigital2025,mcalpineManticoreProjectII2026}, provides a powerful framework for accessing the full cosmological information content. This, however, requires highly accurate forward modelling of both gravitational evolution \citep[e.g.][]{stopyraAccurateFieldlevelInference2024} and the galaxy--matter connection \citep[e.g.][]{Bartlett2024a}. Parameterized models, such as neural networks, provide computationally efficient and flexible representations of complex, non-linear, non-local mappings while remaining differentiable, as required by gradient-based field-level inference. In this work, we introduce a framework that extends field-level inference to incorporate parameterized models without pre-training. Instead, their parameters are inferred directly from the data and guided by physics-informed priors derived from a reference set of simulations that need not be differentiable. \\
\indent
The development of this framework is motivated by a central modelling challenge in cosmological inference: the relation between the underlying matter distribution and the observed galaxies, commonly referred to as \emph{galaxy bias}. Historically, galaxy bias has been modelled in progressively more sophisticated ways. In its simplest form, motivated by the Press–Schechter formalism \citep[e.g.][]{press_schechter74, ps2} and peak-background split arguments \citep[e.g.][]{og_kaiser,peak_background_split}, galaxy bias can be approximated as linear on large scales. Extensions to quasi-linear regimes led to the local bias expansion formalism \citep{local_bias_expansion}, and perturbative bias models now form the basis of many analyses using two- and three-point statistics \citep[e.g.][]{tegmarkCosmologicalParametersSDSS2004,sanchezCosmologicalParameterConstraints2009,ivanovCosmologicalParametersBOSS2020,rizzoHaloBispectrumMultipoles2023a}.

In the context of forward modelling within field-level inference, bias prescriptions are often chosen for their flexibility, differentiability, and computational efficiency, commonly relying on truncated perturbative expansions or empirical local-in-matter-density schemes \citep{neyrinckHaloBiasFunction2014,reppTwostateModelGalaxy2020}. In reality, the galaxy--matter connection is inherently non-linear, non-local, and shaped by temporal processes such as galaxy mergers and environmental history. As increasingly small-scale and higher-order information is incorporated, the simplified prescriptions mentioned above are not sufficient to accurately capture the full information content of the data \citep[e.g.][]{Bartlett2024a}.

To address these limitations, more expressive parametrizations, including neural networks, have been explored \citep[][]{modiCosmologicalReconstructionGalaxy2018,zhangDarkMatterGalaxies2019,pirasFastRealisticLargescale2023,pellejeroibanezHybridBiasDisplacement2024,Ding2024,pandeyCHARMCreatingHalos2024a,Pandey2024a}. A common strategy when employing neural networks in large-scale structure modelling is to train them on ensembles of numerical simulations and subsequently use them as fixed surrogate models of the underlying physical process. Considerable effort is currently devoted to validating such approaches on held-out simulated data sets to assess how accurately they recover the relevant physical mappings. However, inference is ultimately performed on a single realization, the observed Universe, motivating approaches in which simulation-derived physical information can guide the inference without fixing the mapping entirely.

An alternative is therefore to treat the parameters of a parameterized model, such as a neural network, as part of the inference problem itself, allowing them to be constrained directly by the data. The resulting flexibility, however, introduces many additional degrees of freedom and can exacerbate parameter degeneracies in Bayesian inference. Moreover, overparameterized models like neural networks are typically trained on large data ensembles, whereas we have only one Universe, limiting the statistical power available to constrain such models. Even when simulations are not used as training data, they can still be leveraged to construct physics-informed priors that guide the inference towards physically plausible mappings. In cosmological parameter inference studies based on explicit likelihoods, recent works \citep[e.g.][]{ivanovFullshapeAnalysisSimulationbased2024,ivanovFullshapeAnalysisSimulationbased2024a,cabassBOSSConstraintsMassive2024,zhangHODinformedPriorEFTbased2025a,akitsuMappingGalaxyhaloConnection2024,chudaykinReanalyzingDESIDR12026,mooreInformativePriorsPrimordial2026} have demonstrated how simulations can be harnessed to construct such informative priors.

In this work, we extend the concept of simulation-based priors to cosmological field-level inference and validate the framework through a controlled proof-of-concept study using simulated data. We introduce \robin\ (\textbf{RO}bust \textbf{B}ayesian \textbf{IN}ference with \textbf{P}hysics-\textbf{I}nformed \textbf{P}riors), a general and model-agnostic framework that enables joint inference of flexible model parameters and physical parameters under simulation-informed regularization. The method can be implemented as a modular addition to existing inference pipelines and is demonstrated here in the context of galaxy bias modelling. In our demonstration, the physics-informed prior excludes a degenerate bias solution, reduces posterior uncertainties on the bias parameters by up to $20\%$, and reduces the autocorrelation length by approximately $40$--$45\%$, while preserving an unbiased recovery of the initial conditions.

The paper is organized as follows. In \cref{sec:robin_pip}, we review Bayesian inference with an explicit likelihood and discuss the challenges associated with specifying physically meaningful priors before introducing the general \robin\ formalism. In \cref{sec:borg,sec:borg_robin}, we describe the proof-of-concept implementation used in this study. The main results are presented in \cref{sec:results}, and we discuss implications and future directions in \cref{sec:discussion}, before concluding in \cref{sec:conclusion}.
\section{Physics-informed priors from simulations}
\label{sec:robin_pip}

In the context of Bayesian inference, the general aim is to infer a set of parameters $\btheta$ (e.g. cosmological parameters, cosmic initial conditions, or astrophysical parameters) for a given data set $\data$ (e.g. galaxy clustering survey).
The resulting posterior distribution $\calP(\btheta|\data)$ that describes the parameters given the data can then be recovered by applying Bayes' theorem
\begin{equation}
    \calP(\btheta|\data) = \frac{\calP(\data|\btheta)\,\calP(\btheta)}{\calP(\data)},
    \label{eq:bayes}
\end{equation}
where we call $\calP(\data|\btheta)$ the likelihood, $\calP(\btheta)$ the prior, and $\calP(\data)$ the evidence.
In general, the parameters of interest are not directly observable, hence a forward model $\mathcal{M}$ is needed to map them to a predicted data realization $\data' \equiv \mathcal{M}(\btheta)$, which is compared to the observed data $\data$. We can then solve for the posterior using methods such as variational approaches \citep[e.g.][]{hoffmanStochasticVariationalInference2013,knollmullerMetricGaussianVariational2020, frankGeometricVariationalInference2021}, or Markov chain Monte Carlo (MCMC) algorithms \citep[e.g.][]{geyerPracticalMarkovChain1992,speagleConceptualIntroductionMarkov2020}.

For idealized inference tasks, one can construct a deterministic function that accurately describes the possible data space using only the parameters of interest, and then infer the posterior using one of the methods mentioned. However, in most cases, it is necessary to include an additional set of parameters, commonly referred to as \textit{nuisance parameters}. Such parameters account for, e.g., unknown systematics or modelling steps for which the correct parametrization is non-trivial. Inferring nuisance parameters jointly with the parameters of interest, often referred to as self-calibration, can mitigate biases arising from model misspecification by enhancing the forward model's flexibility. In turn, this additional flexibility may reduce the constraining power of the data on the parameters of interest.

\subsection{Challenges in choosing priors for Bayesian inference}

While Bayes' theorem (\cref{eq:bayes}) establishes a framework for principled reasoning, it provides no prescription for constructing the prior $\calP(\btheta)$.
The challenge of prior specification depends on whether physically motivated priors can be constructed for a given set of parameters.
For some parameters, informative priors can be derived from theoretical expectations or external data sets; for instance, cosmological parameters can be constrained using \textit{Planck} results \citep{damicoCosmologicalAnalysisSDSS2020,ivanovCosmologicalParametersBOSS2020} or Big Bang nucleosynthesis priors.
For other parameters, such as those describing complex physical processes or model components with many degrees of freedom, no such external constraints exist.
In these cases, the prior is commonly taken to be weakly informative, such as bounded uniforms or broad Gaussians, which impose minimal structure and let the likelihood dominate, but this leaves the inference vulnerable to degeneracies and unphysical solutions. Another strategy is to adopt simulation-based priors, using external simulations to estimate the distribution of the relevant parameters \citep[e.g.][]{ivanovFullshapeAnalysisSimulationbased2024,ivanovFullshapeAnalysisSimulationbased2024a,cabassBOSSConstraintsMassive2024,zhangHODinformedPriorEFTbased2025a,akitsuMappingGalaxyhaloConnection2024}. For example, \citet{ivanovFullshapeAnalysisSimulationbased2024} applied a neural network to estimate the prior distribution of effective field theory (EFT) bias parameters from a suite of mock catalogues. 

The challenge of prior specification is particularly acute for parameterized models in which individual parameters have no direct physical meaning and cannot be bounded by hand.
A prominent example is a neural network used as part of the forward model: with weights $\boldsymbol{\alpha} \in \mathbb{R}^d$ and $d \gg 10$, no single parameter corresponds to an interpretable physical quantity, making physically motivated prior bounds impossible to specify.

One solution is to exclude such parameters from the probabilistic model entirely by pre-training the network on simulations and keeping the weights fixed during inference \citep[e.g.][]{Doeser2024}.
This sidesteps the prior problem but risks biasing the inference if the training data does not fully represent the target observations or unless another component of the forward model is flexible enough to absorb any residual model misspecification.
Another approach would be to infer the weights jointly with the target parameters, as in Bayesian neural networks \citep[see e.g. the review by][]{goanBayesianNeuralNetworks2020} and used in e.g. \citet{lemosRobustSimulationBasedInference2023}, allowing complex mappings to be learned directly from the data without fixing the model a priori.

In this work, we propose an approach to setting priors on the parameters of any parameterized model when physically motivated bounds cannot be specified by hand.
We present \robin (\textbf{RO}bust \textbf{B}ayesian \textbf{IN}ference with \textbf{P}hysics-\textbf{I}nformed \textbf{P}riors), a simulation-based prior approach for field-level inference that guides the inference towards physically consistent solutions, agnostic to the specific parametrization or dimensionality of the model.

\subsection{Introducing physical constraints}

In Bayesian Origin Reconstruction from Galaxies (\borg; \cref{sec:borg}), the forward model decomposes as $\calM = B_{\boldsymbol{\alpha}} \circ \mathcal{S}$, where the deterministic gravity solver $\mathcal{S}$ maps the cosmic initial conditions $\boldsymbol{x}$ to the dark matter overdensity field $\odm$, and the bias model $B_{\boldsymbol{\alpha}}$ maps $\odm$ to the predicted galaxy field via a parameter vector $\boldsymbol{\alpha}$.
The joint parameter vector becomes
\begin{equation}
    \btheta = \{\boldsymbol{x},\, \boldsymbol{\alpha}\}.
    \label{eq:theta_def}
\end{equation}
Because $\boldsymbol{x}$ and $\boldsymbol{\alpha}$ are sampled jointly, the data likelihood alone cannot disentangle their individual contributions: a combination of incorrect initial conditions statistics and a compensating bias parametrization can yield a galaxy field that is statistically consistent with the data, while the bias mapping $B_{\boldsymbol{\alpha}}$ itself is physically implausible.
To prevent such degenerate solutions, we propose \robin as a framework that imposes simulation-informed priors directly on $B_{\boldsymbol{\alpha}}$, ensuring the bias mapping is consistent with physical expectations from simulations.

Because the constraint targets the output of $B_{\boldsymbol{\alpha}}$ rather than its parameters, \robin is model-agnostic, i.e.\ it does not depend on the specific parametrization of the bias model.
Furthermore, because the simulation ensemble enters only as a prior, not as part of the differentiable forward model, \robin enables the incorporation of non-differentiable physics into gradient-based inference via the simulation-ensemble prior. However, we note that the role of this prior is not necessarily to introduce physical processes absent from the forward model itself, but to restrict the broad family of mappings permitted by a flexible parametrization to those consistent with relations observed in the simulations.

\begin{figure*}[t]
  \centering
  \includegraphics[width=\linewidth]{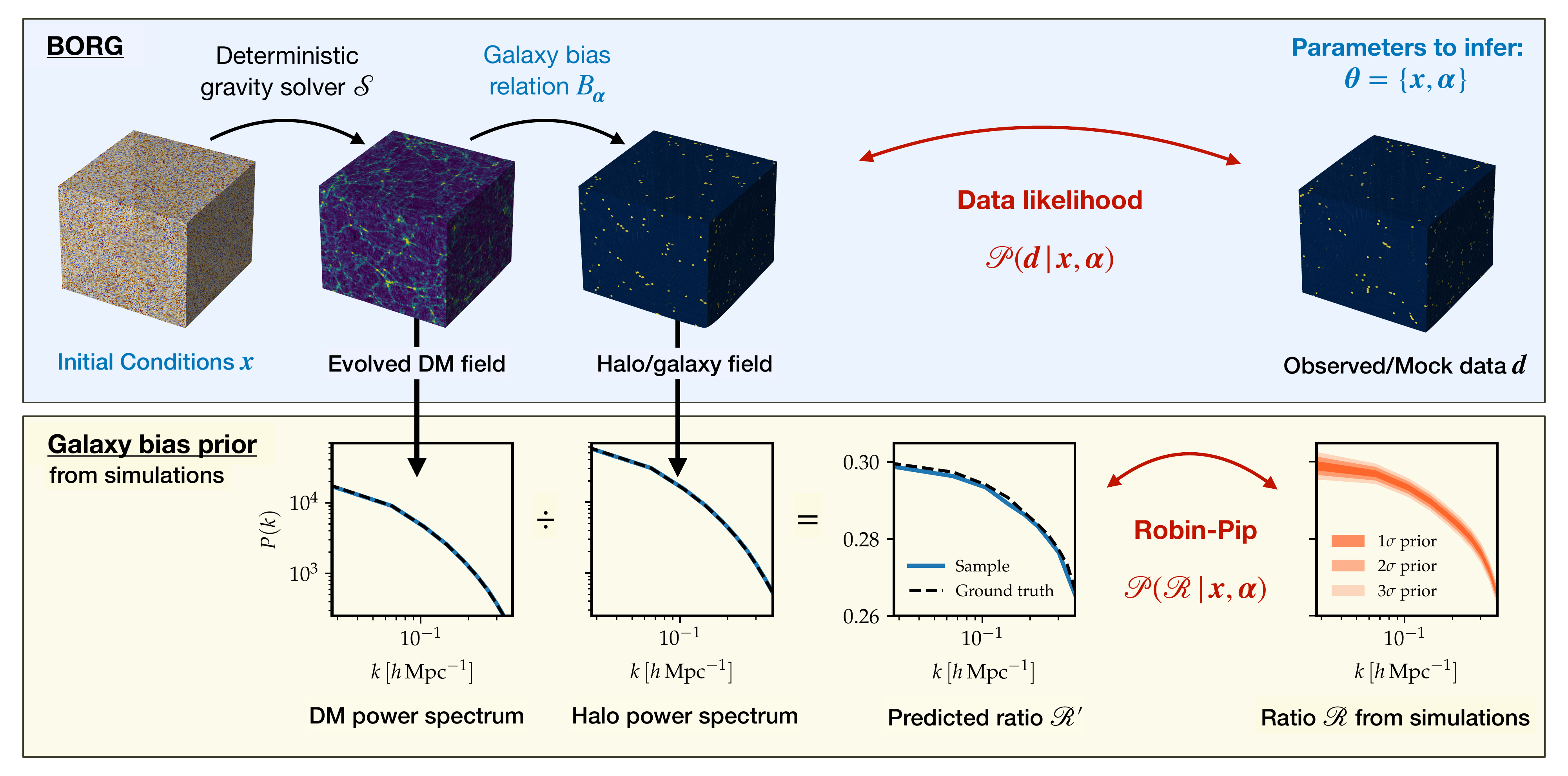}
  \caption{Overview of the \borg inference pipeline augmented with \robin.
  \textit{Top:} The \borg forward model maps the initial conditions $\boldsymbol{x}$ through the deterministic gravity solver $\mathcal{S}$ to the evolved dark matter field, which is converted to the halo/galaxy field via the galaxy bias relation $B_{\boldsymbol{\alpha}}$.
  The data likelihood $\calP(\boldsymbol{d}|\boldsymbol{x},\boldsymbol{\alpha})$ compares the predicted field to the observations $\boldsymbol{d}$ to infer the parameters $\boldsymbol{\theta} = \{\boldsymbol{x}, \boldsymbol{\alpha}\}$.
  \textit{Bottom:} \robin augments the inference by computing the ratio $\sety'(k) = P^{\rm m}(k)/P^{\rm h}(k)$ of the matter and halo power spectra at each iteration and comparing it to the distribution of $\mathcal{R}$ estimated from a reference simulation ensemble (shaded bands indicate $1\sigma$, $2\sigma$, and $3\sigma$).
  This defines the physics-informed prior, implemented in practice as an additional likelihood term $\calP(\mathcal{R}|\boldsymbol{x},\boldsymbol{\alpha})$, that guides the bias parameters $\boldsymbol{\alpha}$ to physically consistent values.}
  \label{fig:general_flow}
\end{figure*}

Concretely, \robin computes a summary statistic $\mathcal{R}$ from intermediate fields of the running forward model and compares it to the same statistic precomputed from a reference simulation ensemble. In principle, $\mathcal{R}$ can be any compressible quantity, from a single power spectrum to the full $N$-point correlation hierarchy. Notably, $\mathcal{R}$ is a function of the combined effect of the model parameters, so the constraint acts on parameter combinations rather than on individual parameters, making it applicable even when no single parameter has a direct physical interpretation.
In this work, we demonstrate \robin using the power spectrum ratio
\begin{equation}
    \mathcal{R}(k) = \frac{P^{\rm m}(k)}{P^{\rm h}(k)}
    \label{eq:power_ratio}
\end{equation}
between the dark matter and halo overdensity fields (see \cref{fig:general_flow}), though the choice of summary statistic is not a fundamental constraint of the framework.

Formally, extending \cref{eq:bayes} with the additional constraint from the simulation ensemble, the augmented inference reads:
\begin{equation}
\calP(\btheta|\data, \sety) = \frac{\calP(\data,\sety|\btheta)\calP(\btheta)}{\calP(\data,\sety)},
\end{equation}
where the simulation ensemble provides the additional conditioning information. Since this simulation-derived information is obtained independently of the observations $\data$, we treat the two sources of information as conditionally independent given $\btheta$, such that $\calP(\data,\sety|\btheta)
= \calP(\data|\btheta)\calP(\sety|\btheta)$.
Here, $\calP(\sety|\btheta)$ quantifies the compatibility of the summary statistic predicted by the current inference state $\btheta$ with the distribution estimated from the simulation ensemble, and not the likelihood of the simulation ensemble itself.

Applying the product rule together with this independence, and then applying Bayes' theorem to the simulation term, gives:
\begin{equation}
	\calP(\btheta|\data, \sety) = \frac{\calP(\data|\btheta)\,\calP(\btheta)}{\calP(\data)} \frac{\calP(\sety|\btheta)}{\calP(\sety)} = \frac{\calP(\data|\btheta)\,\calP(\btheta|\sety)}{\calP(\data)}. \label{eq:bayes_modified_prior}
\end{equation}
The only difference from the original posterior \cref{eq:bayes} is therefore the modified prior $\calP(\btheta|\sety)$, which encodes all simulation-derived information available before analysing the data.

The physics-informed prior from simulations can also be interpreted as an additional likelihood term evaluated against the simulation ensemble rather than the data.
This is evident from the first equality in \cref{eq:bayes_modified_prior}; omitting normalization terms that do not depend on $\btheta$, the posterior is proportional to
\begin{equation}
    \calP(\btheta|\data, \sety) \propto \calP(\data|\btheta)\,\calP(\sety|\btheta)\,\calP(\btheta), \label{eq:bayes_proportional}
\end{equation}
where the three factors are the data likelihood, the simulation constraint, and the prior on $\btheta$, respectively. The resulting total log-likelihood reads
\begin{equation}
	\mathcal{L} = \mathcal{L}_{\boldsymbol{d}} + \mathcal{L}_{\sety} \equiv \log \calP(\data|\btheta) + \log \calP(\sety|\btheta). \label{eq:final_log_lh}
\end{equation}
Through the $\calP(\sety|\btheta)$ term, parameter combinations whose intermediate model outputs fall outside the range expected from simulations are penalized, guiding the inference towards physically plausible solutions without modifying the data likelihood. In \cref{sec:borg_robin} we introduce how this is implemented in practice within \borg.

\section{Field-level inference with \borg}
\label{sec:borg}

To perform field-level inference, we use the Bayesian Origin Reconstruction from Galaxies (\borg; \citealt{jascheBayesianPhysicalReconstruction2013a, jaschePresentCosmicStructure2015, lavauxUnmaskingMaskedUniverse2016, jaschePhysicalBayesianModelling2019, lavauxSystematicfreeInferenceCosmic2019b, mcalpineManticoreProjectDigital2025}) algorithm, introduced in \cref{subsec:borg_overview}. We then describe the specific forward model (\cref{subsec:borg_forward}), challenges with galaxy biasing (\cref{subsec:bias_challenges}), and the sampling strategy (\cref{subsec:hmc}).

\subsection{The \borg framework}
\label{subsec:borg_overview}

\borg is a modular field-level Bayesian inference framework designed to infer the high-dimensional space of plausible cosmic initial conditions that gave rise to the large-scale structure (LSS) observed today.
It has been further developed for joint sampling with nuisance parameters such as galaxy bias, cosmological parameters, and other model components (see \cref{subsec:hmc}); for a full description we refer to \citet{jaschePhysicalBayesianModelling2019}.

Rather than inferring the present-day matter distribution directly, whose non-Gaussian probability distribution makes setting a well-motivated prior challenging, \borg infers the primordial initial conditions.
These are nearly Gaussian, as predicted by inflation and confirmed observationally \citep[e.g.][]{komatsuSEVENYEARWILKINSONMICROWAVE2011, aghanimPlanck2018Results2018}, which admits a well-motivated Gaussian prior \citep[first discussed in][]{jascheBayesianPhysicalReconstruction2013a}.
Furthermore, LSS survey data only provides limited information on the final state of structure formation (due to survey masks, selection effects, and noise), making unique backward integration to the initial conditions infeasible \citep[see e.g.][]{Nusser92,jascheBayesianPhysicalReconstruction2013a}.
\borg therefore never relies on the inversion of the flow of time but always on evaluations of the forward model.
This requires a differentiable forward model that maps the initial conditions through gravitational structure formation all the way to the observed galaxy field, so that gradients of the log-posterior can be propagated back through the full model chain, enabling efficient sampling of the high-dimensional space of plausible initial conditions.

\subsection{Forward model setup}
\label{subsec:borg_forward}

We model the gravitational evolution of the initial matter density field to the late-time dark matter overdensity using the second-order Lagrangian Perturbation Theory (\texttt{2LPT}) solver in \borg.
The 2LPT approximation provides a good balance of accuracy and computational efficiency on the large scales resolved by our simulation box (a periodic cube of side length $1000\,\Mpch$ discretized into $128^3$ voxels), making it well suited for the high-dimensional sampling in this proof-of-concept.
The initial power spectrum is generated following \citet{Eisenstein1999} with cosmological parameters fixed to the fiducial Quijote cosmology \citep{Villaescusa-Navarro2020}, i.e. with parameters $\Omega_\mathrm{m} = 0.3175$, $\Omega_\mathrm{b} = 0.049$, $h = 0.6711$, $n_\mathrm{s} = 0.9624$, $\sigma_8 = 0.834$, and $w=-1$, closely consistent with the Planck Collaboration constraints \citep{aghanimPlanck2018Results2018}.

The evolved matter field is mapped to halo counts via the truncated power-law bias model \citep[henceforth TPL;][]{neyrinckHaloBiasFunction2014},
\begin{equation}
    \lambda_i = \bar{N} (\delta^{\rm m}_i + 1)^\alpha \exp \left[- \left(\frac{\delta^{\rm m}_i + 1}{\rho_{\exp}}\right)^{-\varepsilon}\right],
    \label{eq:truncated_power_law}
\end{equation}
where $\delta^{\rm m}_i$ is the matter overdensity field, $\lambda_i$ the expected halo count in voxel $i$, and $\boldsymbol{\alpha} = \{\bar{N},\, \alpha,\, \varepsilon,\, \rho_{\exp}\}$ the four bias parameters. Here, $\bar{N}$ sets the overall normalization of the expected tracer counts, $\alpha$ controls how strongly the tracer abundance increases with matter density, while $\rho_{\exp}$ sets the characteristic density below which tracer counts are exponentially suppressed and $\varepsilon$ controls the sharpness of this suppression. The choice and limitations of this model are discussed in \cref{subsec:bias_challenges}. 

Finally, the data $\boldsymbol{d} \equiv \boldsymbol{N}^\mathrm{h}$ are the observed tracer counts on the discretized grid, where $N^\mathrm{h}_i$ denotes the count in the $i$-th voxel. The standard likelihood for galaxy clustering \citep{jaschePhysicalBayesianModelling2019} is the Poisson likelihood
\begin{equation}
	\calP(\boldsymbol{d}\,|\,\boldsymbol{x},\boldsymbol{\alpha}) = \prod_{i} \frac{\lambda_i^{N^\mathrm{h}_i}\,e^{-\lambda_i}}{N^\mathrm{h}_i!}.
	\label{eq:lh_poisson}
\end{equation}
In this work, $\boldsymbol{N}^\mathrm{h}$ corresponds to mock halo counts (hence the superscript $h$, rather than the commonly used $g$ for galaxies) in real space; redshift-space distortions, survey masks, and observational systematics are therefore omitted, though all are implemented in \borg for real-data applications \citep[see e.g.][]{jaschePhysicalBayesianModelling2019,mcalpineManticoreProjectDigital2025,mcalpineManticoreProjectII2026}.

\subsection{Challenges with galaxy bias modelling}
\label{subsec:bias_challenges}
All bias prescriptions currently available in \borg are local-in-matter-density (LIMD) models that predict tracer counts based solely on the local matter density, without accounting for the formation history or non-local environmental effects of dark matter haloes. Extensions to more physically motivated prescriptions have been incorporated into \borg analyses \citep{mcalpineManticoreProjectDigital2025, mcalpineManticoreProjectII2026}, but these remain LIMD and do not resolve the fundamental non-locality problem.
More expressive models, including neural networks \citep[e.g.][]{Ding2024,pandeyCHARMCreatingHalos2024a}, offer a promising route towards capturing increasingly non-linear galaxy bias.

When its bias parameters are inferred jointly with the initial conditions, the four-parameter TPL model used here already admits degenerate, unphysical solutions in the absence of informative priors, including solutions with inverted initial conditions and inverted bias relation, as we demonstrate in \cref{sec:results}. Keeping the demonstration at this deliberately simple scale, where ground-truth parameter values are known, allows the effect of \robin\ to be rigorously benchmarked before extending it in future work to the overparameterized neural-network models for which it is ultimately intended.

\subsection{Efficient high-dimensional sampling strategy}
\label{subsec:hmc}

Exploring the space of plausible initial conditions from which the observed three-dimensional large-scale structure of our Universe originated is a significant high-dimensional problem. It entails obtaining a posterior distribution over initial density fluctuations in all regions of space, discretized into a grid of $128^3$ voxels in this work and reaching $256^3 \sim 10^7$ or higher in data applications \citep{jaschePhysicalBayesianModelling2019, mcalpineManticoreProjectDigital2025,mcalpineManticoreProjectII2026}.
Given a multi-dimensional probability distribution $\mathcal{P}(\boldsymbol{x})$, MCMC methods can efficiently sample from it by generating a Markov chain with the desired distribution as its equilibrium distribution. In particular, the Hamiltonian Monte Carlo (HMC) method, based on the Metropolis--Hastings algorithm, draws on concepts from classical physics and Hamiltonian dynamics to efficiently explore the parameter space \citep{Duane1987,nealProbabilisticInferenceUsing1993}; HMC sampling for LSS data was introduced by \citet{Jasche2010a}. To effectively traverse the parameter space with a high acceptance rate and avoid random walk behaviour, HMC utilizes Hamiltonian dynamics, which conserve the Hamiltonian $\mathcal{H}$, constructed from the negative log-posterior and an auxiliary kinetic-energy term, as well as the forward model gradients, and hence relies on differentiable models; we refer to \citet{jascheBayesianPhysicalReconstruction2013a} for details of the HMC implementation used in \texttt{BORG}.

In addition, \borg implements the Gibbs sampling strategy \citep[][]{gemanStochasticRelaxationGibbs1984} to enable joint inference of additional quantities of interest and nuisance parameters, such as the luminosity-dependent galaxy biases \citep{jaschePhysicalBayesianModelling2019,mcalpineManticoreProjectDigital2025,mcalpineManticoreProjectII2026}, galaxy photometric redshift probability density function \citep{Tsaprazi2021}, primordial non-Gaussianities \citep{Andrews2022,Andrews2024}, and cosmological parameters \citep{Ramanah2018,Porqueres2023}.
In this work, we keep the cosmological parameters fixed and focus on jointly sampling the initial conditions $\boldsymbol{x}$ and halo bias parameters $\boldsymbol{\alpha}$ given the data $\boldsymbol{d}$ and the simulation-based prior $\sety$,
\begin{align}
    \icparam &\curvearrowleft \calP(\icparam|\,\biasparam,\data,\sety), \\
    \biasparam & \curvearrowleft \calP(\biasparam|\,\icparam,\data,\sety).
\end{align}
Within this Gibbs scheme, the initial conditions are sampled via HMC, while the bias parameters $\boldsymbol{\alpha}$ are sampled using a multivariate slice sampler \citep{liLatentSliceSampling2020a}, which is well suited to the low-dimensional bias parameter space and avoids the gradient computations and step-size tuning required by HMC.

\section{Guiding bias modelling in \borg with \robin}
\label{sec:borg_robin}

Having introduced the \robin formalism in \cref{sec:robin_pip} and the \borg inference framework in \cref{sec:borg}, this section describes how the two are combined.
We present the implementation of \robin as an augmentation of \borg for joint inference of the initial conditions and galaxy bias parameters, before detailing the implementation (\cref{subsec:robin_extending_borg}), the reference simulation ensemble (\cref{subsec:robin_sim_ensemble}), and the inference configuration (\cref{subsec:robin_inference_config}).
To cleanly isolate the effect of \robin, both the reference ensemble and the mock data are generated with the same \borg forward model used during inference. This controlled setup allows us to test whether the simulation-derived information constrains the space of bias mappings permitted without biasing the inference. Incorporating independent, high-resolution \nbody simulations is deferred to future work, and we discuss this complementary question in \cref{sec:model_misspec}.

\subsection{Extension of the \borg pipeline}
\label{subsec:robin_extending_borg}

\robin is applied by augmenting the \borg likelihood with a simulation-informed prior on the power spectrum ratio $\mathcal{R}$ between the dark matter field and halo field from \cref{eq:power_ratio}.
The ratio estimated from the reference simulation ensemble is denoted $\sety$ (see \cref{subsec:robin_sim_ensemble}), and the ratio predicted by the \borg forward model at each inference step is $\sety'$. 
The high-level workflow is illustrated in \cref{fig:general_flow}.

With \cref{eq:power_ratio}, we can now choose a similarity measure to implement $\log\calP(\sety|\btheta)$ from \cref{eq:final_log_lh} with $\btheta = \{\boldsymbol{x}, \boldsymbol{\alpha}\}$.
We choose a Gaussian distribution to compare the power spectrum ratios, yielding
\begin{equation}
	\log\calP(\sety|\btheta) = -\frac{1}{2}\sum_{k}\frac{\left(\mu_k - \mathcal{R}'_{k}\right)^2}{\sigma^2_{k}},
    \label{eq:robin_logprior}
\end{equation}
where $\mathcal{R}'_k$ is the power spectrum ratio of \cref{eq:power_ratio} evaluated at wavenumber bin $k$, and $\mu_k$ and $\sigma_k$ are the mean and standard deviation of $\mathcal{R}_k$ computed over the simulation ensemble.
Inserting the Poissonian galaxy clustering likelihood from \cref{eq:lh_poisson} and the additional prior constraint of \cref{eq:robin_logprior} into \cref{eq:final_log_lh}, the final log-likelihood then reads:
\begin{equation}
    \mathcal{L} \equiv \sum_i (N^h_i \log \lambda_i - \lambda_i) - \frac{1}{2}\sum_{k}\frac{\left(\,\mu_{k} - \mathcal{R}'_{k}\right)^2}{\sigma^2_{k}},
    \label{eq:robin_final_loglike}
\end{equation}
where we have dropped the constant $\log N^h_i!$ term, as it does not depend on the prediction.  Moreover, the power spectrum computation in \cref{eq:power_ratio} is implemented in JAX \citep{jax2018github}, making it fully differentiable and compatible with the HMC sampler (see \cref{subsec:hmc}).

Effectively, the physics-informed prior from \cref{eq:robin_logprior} constrains the phenomenological bias parameter space $\boldsymbol{\alpha}$.
The Gaussian likelihood guides the power spectra ratio generated by the bias model to the expected mean and standard deviation estimated from the set of simulations $\sety$.

\subsection{Physics-informed simulation ensemble}
\label{subsec:robin_sim_ensemble}

We construct the reference ensemble $\sety$ using the same \borg forward model described in \cref{subsec:borg_forward}.
To ensure that the ground-truth bias parameters used both for mock data generation and for the reference simulations are physically realistic, we calibrate them against the high-resolution Quijote \nbody simulations \citep{Villaescusa-Navarro2020}.
Specifically, we fit the TPL function of \cref{eq:truncated_power_law} to dark matter haloes from the Quijote catalogues, identified with the \texttt{ROCKSTAR} halo finder \citep{Behroozi2013}, with virial masses between $10^{13}\,\hMsun$ and $5 \times 10^{13}\,\hMsun$. This mass range corresponds to the typical halo masses of luminous red galaxy samples \citep[e.g.][]{zhaiClusteringLuminousRed2017}. Rather than adopting the maximum-likelihood solution, we use a No-U-Turn Sampler \citep[NUTS;][]{hoffmanNoUTurnSamplerAdaptively2011a} to draw posterior samples for the bias parameters and randomly draw the ground-truth values adopted in this work (listed in \cref{tab:prior_config}) from this posterior. While we do not aim to reproduce the selection of any particular survey, the resulting halo number density, $\bar{n} = 4.13 \times 10^{-4}\,h^3\,\mathrm{Mpc}^{-3}$, is comparable to that expected for spectroscopic galaxy samples such as \textit{Euclid} \citep{euclidcollaboration:blanchardEuclidPreparationVII2020}.

We then construct the reference ensemble by running $\SI{10000}{}$ \borg\ forward-model realizations, varying both the random seed of the initial conditions and the bias parameters, with the latter sampled from the posterior obtained above. For each realization, we compute the power spectrum ratio defined in \cref{eq:power_ratio}, with the power spectra discretized into $10$ wavenumber bins. From this ensemble, we estimate the mean $\mu_k$ and standard deviation $\sigma_k$ in each bin, which are used in \cref{eq:robin_logprior}. The shaded (orange) bands in the bottom panel of \cref{fig:general_flow} illustrate the resulting prior. Since the current implementation treats the bins independently, finer binning constrains the power spectrum ratio towards the simulation-derived mean more locally in $k$, while neglecting the covariance between bins. We therefore adopt $10$ bins to retain a less restrictive prior on the mean relation. As a preliminary exploration, we also considered a finer discretization with $40$ bins and found no indication of bias from this treatment.

\subsection{Inference configuration}
\label{subsec:robin_inference_config}

We consider two configurations of the uniform priors for the TPL bias model: an Unconstrained (UC) and a Constrained (C) case. In the UC configuration, very broad uniform priors are placed on the bias parameters, mimicking a scenario with effectively no prior knowledge, as could arise for the weights of a neural-network forward model or the parameters in EFT-based analyses. The C configuration applies physically motivated prior bounds as commonly employed in \borg analyses (see \cref{tab:prior_config}). While the broad UC prior formally permits $\bar{N}<0$, the observed non-zero tracer counts constrain $\bar{N}$ away from zero, preventing the sampler from reaching negative values. The remaining bias parameters can take negative values without yielding negative Poisson intensities, as discussed further in \cref{subsec:convergence}.

\begin{table}
\centering
\renewcommand{\arraystretch}{1.3}
    \begin{tabular}{c|c|c|c}
    \hline\hline
    \multirow{2}{*}{Bias parameter} & \multirow{2}{*}{Ground truth} & \multicolumn{2}{c}{Parameter prior} \\
    & & unconstrained & constrained \\
    \hline
    $\bar{N}$ & 0.957 & $\left( -10^8 , 10^8 \,\right) $ & $\left( 0, 10^8 \,\right)$ \\
    $\alpha$  & 0.362 & $\left( -10^8, 10^8 \,\right)$ & $\left( 0, 6 \,\right]$ \\
    $\varepsilon$ & 1.710 & $\left( -10^8, 10^8 \,\right)$ & $\left( 0, 3 \,\right]$ \\
    $\rho_{\exp}$ & 1.475 & $\left( -10^8, 10^8 \,\right) $& $\left( 0, 10 \,\right]$ \\[4pt]
    \end{tabular}
    \caption{TPL bias parameter ground-truth values and prior ranges for the two inference configurations. The ground-truth values are drawn from the Quijote-calibrated posterior as described in \cref{subsec:robin_sim_ensemble}.}
    \label{tab:prior_config}
\end{table}

These two configurations are each run both with and without \robin, resulting in four inference chains. For each configuration, the two chains are run on the same mock observables to recover their respective known ground truth; the configurations are summarized in \cref{tab:run_config}.

\begin{table}
\centering
\renewcommand{\arraystretch}{1.3}
    \begin{tabular}{l|c|c}
    \hline\hline
    Run & Prior & \robin \\
    \hline
    UC               & unconstrained & \xmark \\
    UC\,\&\,\robin   & unconstrained & \cmark \\
    C                & constrained   & \xmark \\
    C\,\&\,\robin    & constrained   & \cmark \\[4pt]
    \end{tabular}
    \caption{Overview of the four inference configurations. The prior column refers to the bias parameter prior ranges in \cref{tab:prior_config}. The \robin column indicates whether the augmented likelihood of \cref{eq:robin_final_loglike} is active.}
    \label{tab:run_config}
\end{table}
 \section{Results}
\label{sec:results}

\begin{figure*}[!t]
    \centering
    \includegraphics[width=1\linewidth]{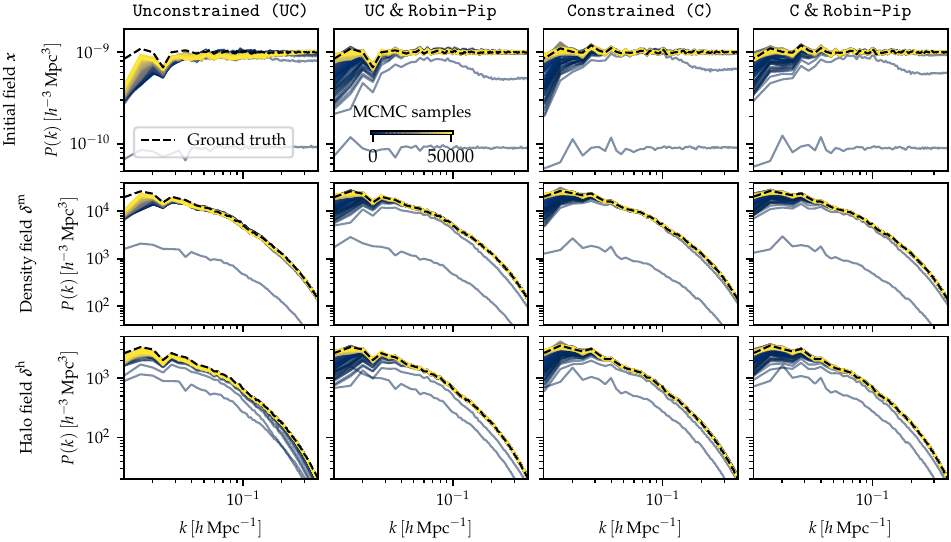}
    \caption{Power spectra of the initial conditions $\boldsymbol{x}$ (top), evolved matter overdensity $\delta^{\mathrm{m}}$ (middle), and halo overdensity $\delta^{\mathrm{h}}$ (bottom) during MCMC warm-up for the four inference configurations. Colours indicate the sample index and dashed black lines the ground truth. The \robin-augmented chains converge towards the ground truth, while UC does not.}
    \label{fig:power_spectra}
\end{figure*}

\begin{figure}[!t]
    \centering
    \includegraphics[width=1\linewidth]{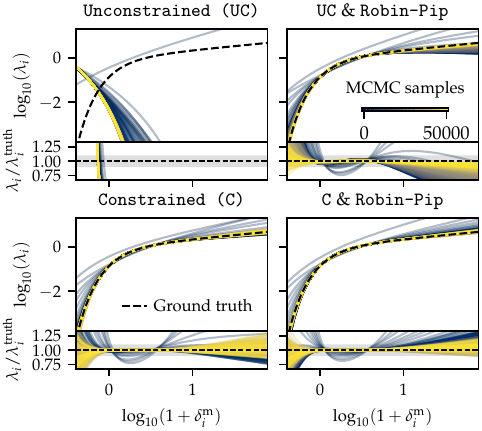}
    \caption{TPL bias function samples $\lambda_i(\delta^{\rm m}_i)$ in voxels $i$ (upper sub-panels) and their ratio to the ground truth (lower sub-panels) for all four inference configurations. Colours indicate the MCMC sample index, dashed black lines the ground truth, and shaded bands a $\pm10\%$ deviation. The UC run (top-left) converges to an inverted bias function, illustrating a degenerate solution, while \robin\ (top-right) recovers the correct relation during warm-up.}
    \label{fig:bias_functions}
\end{figure}

We present results for the four inference configurations defined in \cref{subsec:robin_inference_config}. In all cases, the initial conditions are drawn from an underdispersed Gaussian distribution with variance set to $0.3$ times the cosmological prior variance, providing an initialization far from the target posterior distribution. This provides a stringent test of convergence, since HMC must explore the high-dimensional parameter space from a distant starting point, and mirrors the situation encountered in real data applications where the true posterior is unknown. In \cref{subsec:convergence} we present results for all four configurations; in \cref{subsec:bias_constraints} we provide a more detailed comparison between the constrained chains with and without \robin.

\begin{figure*}[!t]
    \centering
    \includegraphics[width=1\linewidth]{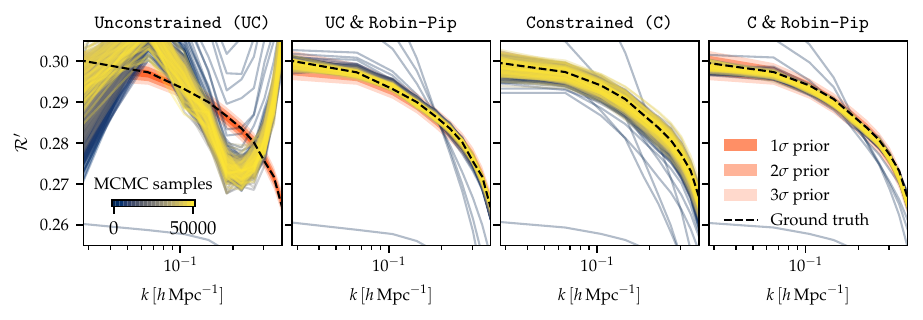}
    \caption{Intermediate power spectrum ratios $\sety'(k) = P^{\rm m}(k) / P^{\rm h}(k)$ over the MCMC chains for the four inference configurations. The shaded orange bands mark the $1\sigma$, $2\sigma$, and $3\sigma$ contours of the \robin prior $\sety$, the dashed black line shows the ground truth, and colours indicate the sample index. Without \robin, the ratios show more scatter for both UC and C, while \robin confines them to the simulation-derived prior.}
    \label{fig:power_ratio_constraint}
\end{figure*}

\begin{figure}[!t]
    \centering
    \includegraphics[width=1\linewidth]{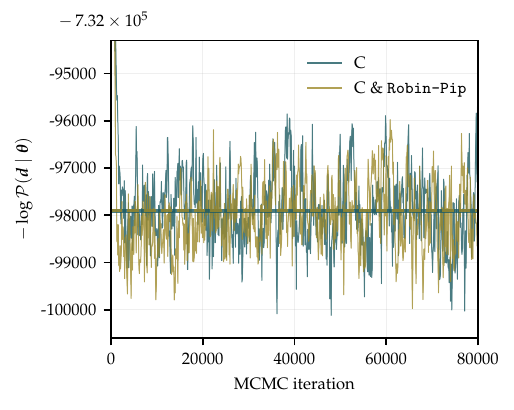}
    \caption{Trace plot of the negative log likelihood $-\mathcal{L}_{\boldsymbol{d}}$ for the halo clustering for the C and C\,\&\,\robin runs. Horizontal lines show the means after iteration $2000$, around which both chains fluctuate without a systematic shift due to \robin.}
    \label{fig:log_lh}
\end{figure}

\subsection{Field power spectra and chain convergence}
\label{subsec:convergence}
To monitor the convergence towards the posterior, \cref{fig:power_spectra} displays the power spectra of a set of HMC samples for the initial conditions field, the evolved matter overdensity field, and the resulting halo overdensity field.
Comparing the first two columns (UC and UC\,\&\,\robin), the \robin-augmented chain converges to the correct large-scale power, whereas the UC run alone fails to do so within the full $\SI{50000}{}$ visualized samples; \robin thus guides the sampler to the physical solution. One may note that it is the large scales that converge slowest. This is the result of the voxel-based Poisson likelihood (\cref{eq:lh_poisson}), which constrains each voxel individually, together with the fact that all voxels are constrained. Large-scale modes, which require coherent, simultaneous adjustments across the entire field, therefore move slowly. The C and C\,\&\,\robin\ chains, shown in the right two columns, reach the target distribution after a few thousand warm-up samples. As can be seen after this warm-up phase, the two chains recover consistent power spectra, with no indication of a systematic bias introduced by \robin. To make this comparison more explicit, Fig.~\ref{fig:app_powspec_ratios} shows the ratios of both the posterior mean and individual posterior samples to the ground truth.

We next examine the bias function in \cref{fig:bias_functions}. The UC chain shown here converged to an inverted initial density field together with a galaxy-bias parametrization characterized by negative $\alpha$ and $\varepsilon$. These negative values reverse the usual behaviour of both terms in the TPL model. In this case, the power-law dependence favours haloes in lower-density regions, while the exponential term suppresses them towards higher densities. Consequently, matter-overdense regions are mapped to voids and underdense regions are populated with haloes, corresponding to the physical inverse of the galaxy-bias relation. This is visible both in the log-scale TPL curves diverging downward from the ground truth (upper sub-panel) and in the ratio $\lambda_i / \lambda_i^{\rm truth}$ deviating substantially from unity (lower sub-panel). Together, the inverted initial density field and inverted bias relation form a degenerate solution to the halo clustering likelihood, reproducing the mock halo data similarly well as the physical solution. Note that, in the single UC inference considered here, the sampler converges to this inverted configuration. This demonstrates that such a degenerate solution is permitted by the clustering likelihood but does not establish how frequently UC inference converges to this solution. The physics-informed prior excludes this solution by constraining the inferred bias relation to remain consistent with the simulation-derived physical information.

This flip is also evident in the cross-correlation coefficient $r_c(k)$ between the inferred and ground-truth fields: the UC run yields $r_c^{\rm initial} \approx -0.75$, whereas the halo field remains positively correlated, a direct signature of the inverted-bias solution.
Importantly, the three non-degenerate configurations produce nearly coincident cross-correlation coefficients throughout the probed range, confirming that \robin does not bias the recovered initial conditions (see \cref{fig:cross_corr} in Appendix~\ref{app:cross_corr}).

Related to the bias function is exactly the power spectrum ratio $\mathcal{R}$ enforced via \robin. To examine how well the ratio is tracked across the four inference configurations, \cref{fig:power_ratio_constraint} shows $\sety'(k)$ over all MCMC samples.
Without \robin, noticeable scatter in $\sety'(k)$ is observed, particularly in the unconstrained case; with \robin, the C\,\&\,\robin samples are confined tightly around the prior $\sety(k)$.
Taken together, the power spectrum ratios (\cref{fig:power_ratio_constraint}), the bias function shapes (\cref{fig:bias_functions}), and the cross-correlation coefficients (\cref{fig:cross_corr}) confirm that \robin steers intermediate model predictions towards physically consistent solutions without biasing the recovered initial conditions.

\subsection{Posterior constraints and sampling efficiency}
\label{subsec:bias_constraints}

We now compare only the two constrained configurations (C and C\,\&\,\robin), which both converge to the ground truth, to allow a fair quantitative comparison.
Adding to the convergence diagnostics in \cref{fig:power_spectra,fig:bias_functions,fig:power_ratio_constraint}, the log-likelihood trace in \cref{fig:log_lh} shows that after a few thousand warm-up samples the likelihood settles into a bounded band, confirming that both chains are sampling from the target posterior and that \robin does not shift the overall likelihood value.
We additionally assess convergence by computing the Gelman--Rubin diagnostic \citep{gelmanBayesianDataAnalysis1995} between two $10\,000$-iteration segments of each chain, separated by $10\,000$ MCMC iterations. The median statistic across all initial condition voxels is $\hat{R}_{\mathrm{GR}}=1.015$ for both C and C\,\&\,\robin, while the median across the bias parameters is $\hat{R}_{\mathrm{GR}}=1.016$ and $1.025$, respectively. All values are close to unity, as expected for converged chains.

\begin{figure}[!t]
    \centering
    \includegraphics[width=1\linewidth]{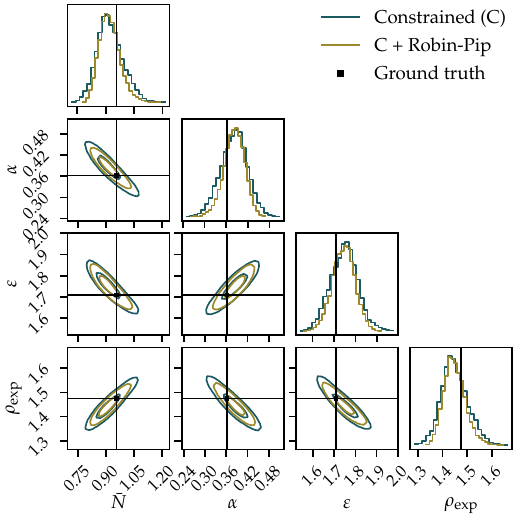}
    \caption{Corner plot of the posterior distributions of the four TPL bias parameters for the constrained inference runs, without (C) and with (C\,\&\,\robin) \robin. One-dimensional marginals are shown along the diagonal, and two-dimensional joint posteriors are shown in the off-diagonal panels; black lines mark the ground-truth values. With \robin, all four parameters are more tightly constrained, with posterior standard deviations reduced by up to $20\%$.}
    \label{fig:bias_corner_constraint}
\end{figure}

\Cref{fig:bias_corner_constraint} shows the marginalized and joint posterior distributions of the four TPL bias parameters for the C and C\,\&\,\robin configurations.
With \robin active, all four parameters exhibit narrower distributions, with posterior standard deviations reduced by up to $20\%$ compared to C alone. The corresponding trace plots of the individual bias parameters are provided in \cref{fig:trace_bias} in Appendix~\ref{app:bias_traces}.

\begin{figure}
    \centering
    \includegraphics[width=1\linewidth]{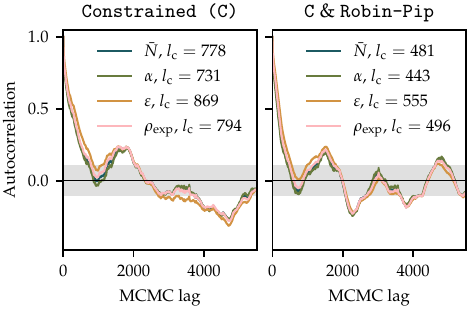}
    \caption{Autocorrelation functions of the four TPL bias parameters as a function of sample lag, for C (left) and C\,\&\,\robin (right). The autocorrelation lengths $l_c$ reported in the legends are reduced by approximately $40$--$45\%$ when \robin is active.}
    \label{fig:bias_acf_constraint}
\end{figure}

The improved sampling efficiency brought by \robin is further quantified by the autocorrelation functions of the four bias parameters in \cref{fig:bias_acf_constraint}. The autocorrelation lengths decrease from $l_c \sim 730$--$870$ samples for C to $l_c \sim 440$--$550$ samples for C\,\&\,\robin, corresponding to a reduction of approximately $40$--$45\%$ across all four parameters. Shorter autocorrelation lengths imply that more statistically independent samples can be obtained, directly increasing the effective sample size.

To further characterize how \robin influences the inference dynamics, we decomposed the total log-likelihood from \cref{eq:robin_final_loglike} into the original clustering contribution $\mathcal{L}_{\boldsymbol{d}}$ and the \robin term $\mathcal{L}_{\sety}$, and computed the mean Euclidean norm of the gradient of each term over the posterior chain:
\begin{equation}
	\text{Avg}_{\,\btheta}({\rm p}) \equiv
        \text{Mean}\left(\left\{\left\lVert \nabla_{\btheta} \mathcal{L}_{\rm p} \right\rVert_i : i = 1, 2, \dots, N_{\rm s}  \right\}\right),
	\label{eq:robin_gradient_norm}
\end{equation}
where $\btheta \in \{\boldsymbol{x}, \boldsymbol{\alpha}\}$ denotes either the initial conditions $\boldsymbol{x}$ or the bias parameters $\boldsymbol{\alpha}$, $\mathrm{p} \in \{\boldsymbol{d}, \mathcal{R}\}$ labels the likelihood term (the data clustering likelihood $\mathcal{L}_{\boldsymbol{d}}$ or the \robin prior $\mathcal{L}_{\sety}$), $i$ is the MCMC sample index running from $1$ to $N_{\rm s}$, the total number of posterior samples, and $\left\lVert\cdot\right\rVert$ is the Euclidean norm.

The results are reported in \cref{tab:robin_gradient_norms}.
For the initial conditions, the \robin gradient contribution ($\text{Avg}_{\boldsymbol{x}}(\sety) = 3.43$) is negligible compared to the clustering likelihood ($\text{Avg}_{\boldsymbol{x}}({\boldsymbol{d}}) \approx 1290$), confirming that \robin does not interfere with the reconstruction of the large-scale structure.
For the bias parameters, however, the \robin gradient ($\text{Avg}_{\boldsymbol{\alpha}}(\sety) = 1843$) is comparable to, and slightly exceeds, the clustering likelihood contribution ($\text{Avg}_{\boldsymbol{\alpha}}({\boldsymbol{d}}) = 1656$), indicating that \robin substantially shapes the bias parameter posterior.
This asymmetry directly reflects the design of \robin: the power spectrum ratio $\mathcal{R}_k$ is sensitive to the bias function through its dependence on $P^{\rm h}_k$, while providing relatively little gradient information for the initial conditions field beyond what the clustering likelihood already captures.

\begin{table}
    \centering
    \renewcommand{\arraystretch}{1.3}
    \begin{tabular}{c|c|c|c}
	    \hline\hline
	    \multirow{2}{*}{param. set $\btheta$ } & Constrained (C) & \multicolumn{2}{c}{C\,\&\,\robin} \\
				   & $\text{Avg}_{\,\btheta}({\boldsymbol{d}})$ & $\text{Avg}_{\,\btheta}({\boldsymbol{d}})$ &  $\text{Avg}_{\,\btheta}(\sety)$ \\
	    \hline
	    $\boldsymbol{x}$ & $1275.10$ & $1290.49$ & $3.43$ \\
	    $\boldsymbol{\alpha}$ & $1588.68$ & $1656.16$ & $1843.04$ \\[4pt]
    \end{tabular}
    \caption{Mean gradient norms over the posterior chain for the individual likelihood contributions ($\mathcal{L}_{\boldsymbol{d}}$ and $\mathcal{L}_{\sety}$) with respect to the initial conditions $\boldsymbol{x}$ and bias parameters $\boldsymbol{\alpha}$, as defined in \cref{eq:robin_gradient_norm}. The \robin gradient dominates for the bias parameters while being negligible for the initial conditions field, confirming that \robin acts mostly on the bias parameter posterior.}
    \label{tab:robin_gradient_norms}
\end{table}

\section{Discussion}
\label{sec:discussion}

The results confirm that \robin guides field-level inference towards physically consistent solutions without biasing the recovered initial conditions or shifting the data likelihood.
In the unconstrained setting, \borg, without prior guidance, can converge to a physically degenerate solution in which anti-correlated initial conditions are paired with an inverted bias mapping that nonetheless reproduces the mock halo data; \robin resolves this by penalizing power spectrum ratios that are inconsistent with the simulation ensemble.
In the constrained setting, \robin further tightens posterior constraints and improves sampling efficiency, demonstrating its benefit even when the inference is already well-behaved, as in the C case without \robin.

\subsection{Towards more expressive forward models}
\borg relies on differentiable forward modelling to exploit gradient information during sampling, restricting the model components that can be incorporated directly into the inference. \robin provides a route to retain this differentiable inference while incorporating physical information from reference simulations that need not themselves be differentiable. Since the resulting simulation-based prior is agnostic to the parametrization of the bias mapping, it can guide substantially more expressive differentiable models towards physically plausible regions of parameter space. Future work will therefore explore the application of \robin to more flexible bias models, such as neural networks \citep[][]{Charnock2020,Ding2024,pandeyCHARMCreatingHalos2024a,Pandey2024a}. Unlike the TPL model considered here, such models can capture non-local environmental effects and operate in substantially larger parameter spaces, where simulation-informed guidance may become increasingly important for constraining otherwise weakly determined degrees of freedom.

Extending the Gibbs sampling scheme to more expressive bias models will, however, require revisiting the sampling strategy. Depending on the architecture, the number of free parameters may range from relatively few when physical symmetries are explicitly encoded \citep[e.g.][]{Charnock2020,Ding2024} to substantially larger numbers of weights in more flexible neural networks \citep[e.g.][]{pandeyCHARMCreatingHalos2024a,Pandey2024a}. Multivariate slice sampling is effective for the low-dimensional bias parameter spaces considered here, but is unlikely to scale to the latter, where gradient-based samplers such as HMC or NUTS \citep{hoffmanNoUTurnSamplerAdaptively2011a} might be more appropriate.

\subsection{High-fidelity simulations and model misspecification}
\label{sec:model_misspec}
A central goal of modern field-level inference is to incorporate increasingly realistic simulations into Bayesian inference. This is being pursued through several complementary directions, including differentiable forward models \citep{Jamieson2024,Jamieson2022,Bartlett2024,Doeser2024,Ding2024,Pandey2024a,horowitzBaryonBridgeStochasticInterpolant2025,Bourdin2024,horowitzJFoFGPUCluster2025}, neural optimisers for non-differentiable simulations \citep{doeserLearningUniverseLearning2025}, and, in this work, \robin, which enables independent physical information extracted from simulation ensembles to guide Bayesian inference. A natural next step is to construct the physics-informed prior from independent, higher-fidelity simulations. For example, the relation between the matter field and haloes or galaxies could be characterized from high-resolution $N$-body or hydrodynamical simulations and used to constrain a flexible bias model within the inference forward model. Since the reference simulations enter only through the prior, they need not be differentiable or share the same forward model used during inference, allowing \robin\ to incorporate physical information from arbitrary simulation pipelines within gradient-based inference.

Using an independent simulation suite does not, by itself, constitute model misspecification. The summary statistic informing the prior can be constructed from the matter field evolved with the inference forward model and the corresponding halo or galaxy distribution from the high-fidelity simulation, thereby constraining the flexible bias mapping towards the relation predicted by the reference simulation. A distinct limitation is that cosmological simulations remain imperfect representations of the real Universe. Even state-of-the-art $N$-body and hydrodynamical simulations \citep[e.g.][]{duboisIntroducingNEWHORIZONSimulation2021,feldmannFIREboxSimulatingGalaxies2023,10.1093/mnras/stac3620,schayeFLAMINGOProjectCosmological2023,weinbergerArepoPublicCode2020a,Springel2021,teyssierCosmologicalHydrodynamicsAdaptive2002} rely on approximations and sub-grid prescriptions, particularly for baryonic processes, leading to differences between them as shown in e.g. \citet{villaescusa-navarroCAMELSProjectCosmology2021,bayerFieldLevelComparisonRobustness2025}. Consequently, the simulations informing the physics-informed prior may not accurately represent the true galaxy--matter relation in the observed Universe. More generally, model misspecification can lead to biased cosmological inference, motivating complementary approaches to diagnose discrepancies between forward models and observations \citep[see, e.g.,][]{hoellingerDiagnosingSystematicEffects2025}. Rather than resolving this limitation directly, \robin\ provides a statistically principled framework for incorporating physical information derived from the best available simulations into Bayesian inference. As physical modelling continues to improve, whether through more accurate simulations or more expressive forward models, these advances can be naturally integrated within \robin.

To reflect the imperfect nature of the simulation-derived prior, \robin\ can be extended in several ways. One possibility is to construct the prior from ensembles generated by multiple simulation suites, thereby propagating modelling uncertainties into the inference. Alternatively, the influence of the prior can be reduced by introducing a hyper-parameter $\beta \leq 1$ into \cref{eq:robin_final_loglike},
\begin{align}
\mathcal{L} &\equiv \log \calP(\data|\btheta) + \beta \log \calP(\sety|\btheta) 
\label{eq:robin_mod_log_lh}
\end{align}
where $\beta<1$ effectively inflates the prior distribution, reducing the influence of the simulation-derived information when its reliability is uncertain. Even with reduced influence, the simulation-derived prior retains its role in guiding an overparameterized model towards physically plausible mappings, allowing its parameters to be jointly inferred with the cosmological fields from the observed Universe.

\subsection{Extending \robin with other summaries and to other parameters}
While the power spectrum ratio used here is a deliberate and conservative starting point, \robin can, in principle, be extended to any compressible statistic, including higher-order correlations or the full $N$-point hierarchy, thereby placing progressively tighter constraints on the forward model.
Richer statistics, however, place greater demands on the reliability of the simulations that define the prior, and, as discussed above, various state-of-the-art simulations exhibit well-documented discrepancies among themselves.
This is precisely why the simulation constraint in \robin is formulated as a Bayesian prior rather than a hard constraint: the model remains free to deviate from the simulation predictions wherever the data likelihood demands it, with the influence of the prior controlled explicitly through $\beta$ in \cref{eq:robin_mod_log_lh}.
This stands in contrast to pre-trained machine learning models, where knowledge from simulations is absorbed into the model weights during training and is therefore not directly updated by the observed data during inference, unless an additional flexible model component is included in the forward model. 

Beyond galaxy bias, \robin can also be used to regularize other parameters. Bayesian self-calibration, which models unknown systematic contamination as nuisance parameters to be sampled, has shown promise in field-level inference \citep[e.g.][]{jascheBayesianPowerSpectrum2017,porqueresExplicitBayesianTreatment2019,lavauxSystematicfreeInferenceCosmic2019b} and weak lensing \citep[e.g.][]{yaoEffectsSelfcalibrationIntrinsic2017,porqueresFieldlevelInferenceCosmic2023}, but comes at the cost of introducing additional nuisance parameters that may themselves suffer from biases. Furthermore, EFT-based forward models have been developed for field-level inference, introducing additional bias and noise parameters that must be constrained jointly with the cosmological fields \citep[see e.g.][]{schmidtUnbiasedCosmologyInference2020,schmidtNthOrderLagrangian2021,kosticConsistencyTestsField2023}. Simulation-based priors have recently been developed to constrain EFT parameters in full-shape analyses \citep[][]{ivanovFullshapeAnalysisSimulationbased2024a,ivanovFullshapeAnalysisSimulationbased2024}, and \robin\ could provide one way towards incorporating analogous simulation-derived constraints directly within EFT-based field-level inference. 

\section{Conclusion}
\label{sec:conclusion}

We have presented \robin, a framework that augments field-level Bayesian inference with physics-informed constraints on intermediate outputs of the forward model, without modifying the data likelihood or requiring knowledge of the bias model's functional form.
Applying \robin within the \borg framework as a proof-of-concept with a four-parameter TPL halo bias model, we showed in an unconstrained configuration that \robin can guide the sampler away from a physically degenerate inverted-bias solution that unconstrained \borg analyses can converge to.
In the constrained configuration, \robin tightened posterior constraints on all four bias parameters by up to $20\%$ and reduced the autocorrelation length by approximately $40$--$45\%$, directly increasing the effective sample size.

In summary, \robin\ provides a model-agnostic framework that allows parameterized forward-model components to be guided by physics-informed priors derived from simulations; the key design choices are the simulation ensemble and summary statistic. While demonstrated here for a four-parameter galaxy bias model, the framework can in principle be applied to other nuisance parameters or forward-model components for which simulation-based expectations can be computed, although its performance in such applications remains to be tested.
These properties make \robin a natural candidate for enabling expressive mappings, such as neural networks, in future cosmological field-level inference, with the potential to fully exploit the information content of non-linear survey data.

\begin{acknowledgements}
	We thank Florent Leclercq and Antoine Gilles--Lordet for their comments on the manuscript. This work has made use of the Infinity cluster hosted by Institut d’Astrophysique de Paris.
	We thank the efforts of S. Rouberol for running the cluster smoothly. This work has utilized the Sunrise HPC facility supported by the Technical Division at the Department of Physics, Stockholm University.
	This work was granted access to the HPC resources of Joliot Curie/TGCC (Très Grand Centre de Calcul) under the allocation AD010413589R1.
	This work was enabled by the research project grant ‘Understanding the Dynamic Universe’ funded by the Knut and Alice Wallenberg Foundation under Dnr KAW 2018.0067.
	JJ acknowledges support from the Swedish Research Council (VR) under the project 2020-05143 -- `Deciphering the Dynamics of Cosmic Structure'.
	GL and SD acknowledge the grant GCEuclid from `Centre National d’Etudes Spatiales' (CNES).
	This work was supported by the Simons Collaboration on `Learning the Universe'. LD also acknowledges financial support from the Simons Foundation through the Center for Computational Astrophysics Pre-doctoral Program at the Flatiron Institute. This work is conducted within the Aquila Consortium (\url{https://aquila-consortium.org}).
\end{acknowledgements}

\section*{Data availability}

The Quijote simulations used in this work are publicly available as described in \citet{Villaescusa-Navarro2020}. The mock data, reference simulation ensemble, and MCMC chains generated for this work are available from the authors upon reasonable request.


\bibliographystyle{aa} 
\bibliography{ROBIN_paper} 



\begin{appendix}
\section{Reconstruction accuracy relative to ground truth}
\label{app:cross_corr}

\begin{figure}[t!]
    \centering
    \includegraphics[width=1\linewidth]{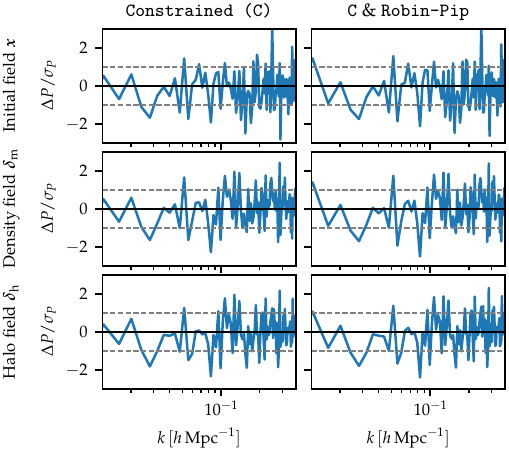}
    \caption{Standardized power-spectrum residuals for the C (left) and C\,\&\,\robin\ (right) configurations after warm-up, for the initial, matter-density, and halo fields (top to bottom). We show $\Delta P/\sigma_P$, where $\Delta P(k) \equiv \langle P(k)\rangle_{\rm post}-P_{\rm truth}(k)$ and $\sigma_P(k)$ is the standard deviation of the posterior power-spectrum samples in each $k$-bin. The dashed lines indicate $\pm1\sigma_P$. The C and C\,\&\,\robin\ residuals fluctuate around zero and show consistent behaviour across scales, demonstrating that the recovered power spectra remain unbiased when including \robin.}
    \label{fig:app_powspec_ratios}
\end{figure}

To complement the power-spectrum evolution shown in Fig.~\ref{fig:power_spectra}, we quantify the agreement with the ground truth after discarding the first $10\,000$ samples in Fig.~\ref{fig:app_powspec_ratios}. We show the difference between the mean power spectrum across posterior samples and the ground-truth power spectrum, normalized by the posterior standard deviation. This allows potential systematic biases to be identified more clearly. For the initial conditions, the matter-density, and the halo fields, the C and C\,\&\,\robin\ residuals are centered around zero and show consistent behaviour across scales, indicating that the inclusion of \robin\ does not bias the recovered power spectra.

\begin{figure*}[t!]
    \centering
    \includegraphics[width=1\linewidth]{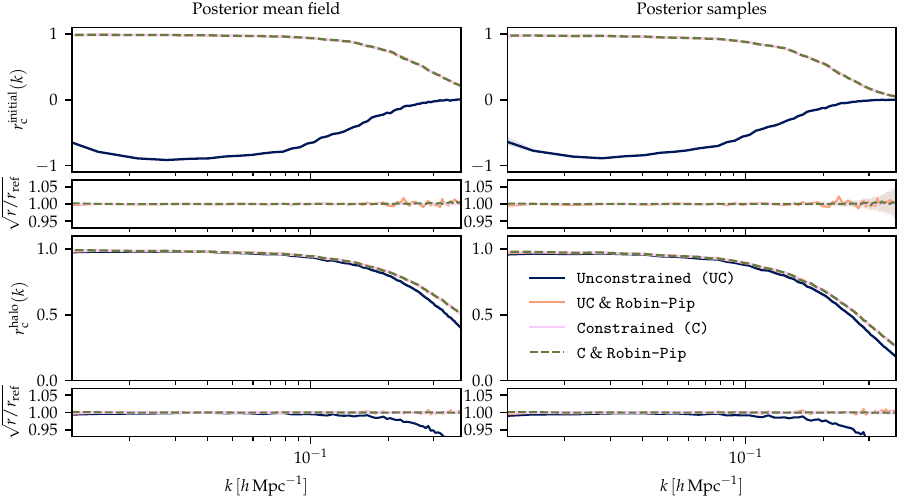}
    \caption{Cross-correlation coefficients $r_c(k)$ between the inferred and ground-truth initial conditions (top) and halo overdensity fields (bottom), for all four inference configurations. The left column shows the cross-correlation of the posterior mean field with the ground truth, while the right column shows the median cross-correlation of individual posterior samples, with shaded bands indicating the 16th--84th percentiles. The smaller sub-panels show the ratio $\sqrt{r_c/r_c^{\rm ref}}$, where we here use the C configuration as the $\mathrm{ref}$. The negative $r_c^{\rm initial}$ in the UC run without \robin\ results from the inverted bias solution shown in \cref{fig:bias_functions}, while the corresponding halo field remains positively correlated with the ground truth, illustrating the degeneracy of this solution. The three non-degenerate configurations produce nearly coincident coefficients, confirming that \robin\ does not bias the recovered fields.}
    \label{fig:cross_corr}
\end{figure*}

Another comparison with the ground truth is provided by the cross-correlation coefficient $r_c(k)$ between two fields $a$ and $b$, defined as
\begin{equation}
    r_c(k) = \frac{P_{ab}(k)}{\sqrt{P_{aa}(k)\,P_{bb}(k)}},
\end{equation}
where $P_{ab}(k)$ is the cross-power spectrum and $P_{aa}(k)$ and $P_{bb}(k)$ are the respective auto-power spectra, so that $r_c=1$ indicates perfect correlation and $r_c=-1$ perfect anti-correlation. \Cref{fig:cross_corr} shows the cross-correlation between the inferred and ground-truth fields, providing a complementary field-level view of the convergence behaviour discussed in \cref{subsec:convergence}.

The top panel shows the cross-correlation with the ground-truth initial conditions and the bottom panel with the ground-truth halo field.
The strongly negative \(r_c^{\rm initial} \approx -0.75\) in the UC run without \robin is a direct fingerprint of the inverted bias solution: anti-correlated initial conditions combined with an inverted bias mapping can reproduce the observed halo field almost as faithfully as the physical solution. That the halo field simultaneously remains positively correlated confirms that the UC run has converged to a degenerate but self-consistent configuration.
For the three non-degenerate configurations the cross-correlation coefficients nearly coincide, remaining above \(r_c \approx 0.75\) until Poisson shot noise begins to dominate at small scales, confirming that \robin does not bias the recovered initial conditions field.

\section{Bias parameter trace plots}
\label{app:bias_traces}

\begin{figure*}[!t]
    \centering
    \includegraphics[width=\linewidth]{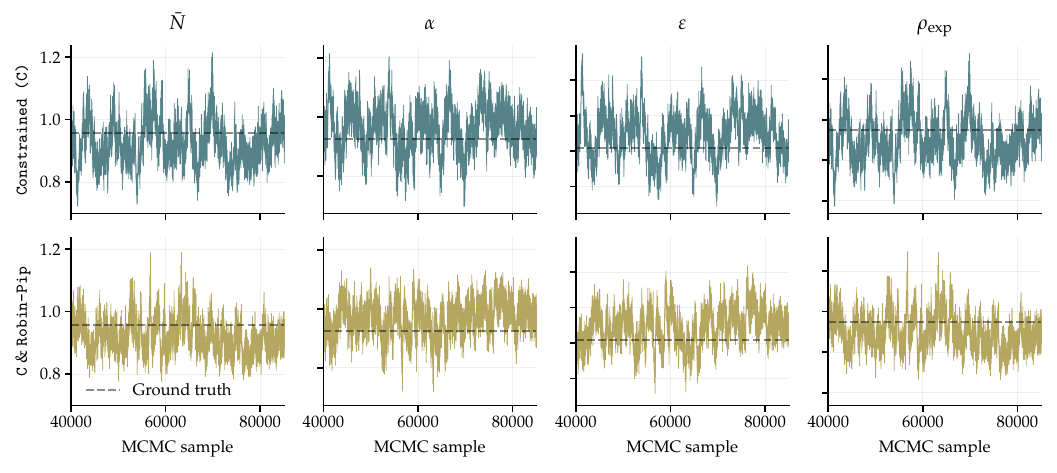}
    \caption{Trace plots of the four TPL bias parameters ($\bar{N}$, $\alpha$, $\varepsilon$, $\rho_{\exp}$) for the two constrained inference configurations. Dashed horizontal lines mark the ground-truth values. The C and C\,\&\,\robin runs both fluctuate around the ground truth, with the \robin-augmented chain exhibiting reduced scatter.}
    \label{fig:trace_bias}
\end{figure*}

\Cref{fig:trace_bias} shows the MCMC trace of each of the four TPL bias parameters (\(\bar{N}\), \(\alpha\), \(\varepsilon\), \(\rho_{\exp}\)) over the last \(\SI{45000}{}\) samples for the two constrained inference configurations.
The traces make the effect of \robin on the sampling dynamics directly visible. The tighter scatter of the C\,\&\,\robin traces around the ground-truth values is consistent with the reduced autocorrelation lengths reported in \cref{fig:bias_acf_constraint}.

\end{appendix}


\end{document}